**Classification:** BIOLOGICAL SCIENCES, Biophysics and Computational Biology

# Balance between cell survival and death: a minimal quantitative model of tumor necrosis factor alpha cytotoxicity

Roberto Chignola[a,b,1], Marcello Farina[c], Alessio Del Fabbro[b],

Edoardo Milotti[b,d]

[a]*Dipartimento di Biotecnologie, Università di Verona, Strada le Grazie 15 - CV1, I-37134, Verona, Italia*
[b]*Istituto Nazionale di Fisica Nucleare, Sezione di Trieste, Via Valerio 2, I-34127, Trieste, Italia*
[c]*Dipartimento di Elettronica e Informazione, Politecnico di Milano, Via Ponzio 34/5, I-20133, Milano, Italia*
[d]*Dipartimento di Fisica, Università di Trieste, Via Valerio 2, I-34127, Trieste, Italia*

[1]**Correspondence to**: Roberto Chignola, Dipartimento di Biotecnologie, Università di Verona, Strada Le Grazie, 15 - CV1, I-37134 Verona, Italia, roberto.chignola@univr.it

**Manuscript Information**: 26 pages, 6 figures

**Abbreviations**: TNF, tumor necrosis factor alpha; TNF-R1, TNF receptor type 1; PLADD, pre-ligand assembly domain; DD, death domain; SODD, silencer of death domain; DISC, death inducing signaling complex

**Abstract**

Tumor Necrosis Factor alpha (TNF) initiates a complex series of biochemical events in the cell upon binding to its type R1 receptor (TNF-R1). Recent experimental work has unravelled the molecular regulation of the recruitment of initial signaling complexes that lead either to cell survival or death. Survival signals are activated by direct binding of TNF to TNF-R1 at the cell membrane whereas apoptotic signals by endocytosed TNF/TNF-R1 complexes. Here we investigate these aspects by developing a quantitative mathematical model of TNF binding, internalization and intracellular signaling. Model outputs compare favorably with experimental data and allow to compute TNF-mediated cytotoxicity as observed in different cell systems. We extensively study the space of parameters to show that the model is structurally stable and robust over a broad range of parameter values. Thus, our model is suitable for implementation in multi-scale simulation programs that are presently under development to study the behavior of large tumor cell populations.

\body

**Introduction**

Tumor Necrosis Factor alpha (TNF) is a cytokine that acts as a key regulator of immune functions, and indeed it exerts pleiotropic effects in immunity, inflammation, control of cell proliferation, differentiation and apoptosis (1-3). Other molecules share with TNF part of its effects and, not suprisingly, they are all evolutionarily related to TNF (4). It is now recognized that TNF is the prototypical member of a growing family of cytokines (4) but, unlike the other members, TNF can trigger intracellular signals that lead either to cell survival and proliferation or death (1-4). This dual role is important as far as the regulation of immune response is concerned, since it acts as a molecular basis for cellular homeostasis. However, TNF is also a natural anti-tumor molecule and these opposing signals might lead to inhibition of tumor growth or, at the opposite, in the promotion of tumor development through direct (see e.g. ref. 5) and indirect mechanisms (e.g. by tissue remodelling and stromal development. See e.g. ref. 6). Dissecting out the molecular mechanisms of TNF signaling is therefore crucial.

Several works have addressed the biochemical pathways of the TNF signaling cascade in a variety of cells and many molecular actors of this complex intracellular machinery have been discovered and studied (for a comprehensive review see ref. 7). The astonishing complexity of the signal transduction machinery triggered by TNF has also attracted the attention of modelers who, using modern approaches common in systems biology, have attempted to unravel the switching mechanism between the pathways leading to cell survival or death through mathematical modeling and theoretical analysis (8-12). These works have been focused on the interplay among intracellular molecules and on the network of reactions stimulated by binding of TNF on its type 1 receptor (TNF-R1). Recent data have clearly shown that the path that leads to cell survival is triggered by TNF binding to its receptor at the cell surface whereas the cell death pathway is triggered by internalized TNF/receptor complexes (13,14).

Here we focus on this important bifurcation between the two paths that occur early upon binding of TNF to TNF-R1, and develop a minimal model of TNF action that is suitable for inclusion in a numerical program that we are currently developing to study tumor spheroids (15,16). The actual feasibility of such a simulation program depends critically on a series of simplification steps, and in particular we proceed in a partly phenomenological way that leads to straightforward parameterizations. The present model follows the same basic principles and replaces some complex molecular pathways with simpler mechanisms, and yet it successfully captures the basic features of experimental toxicity data.

In the next sections, we briefly review current biological knowledge on TNF signaling in cells. Then we present the model and show how model outputs compare with actual data. Finally we explore the parameter space over a wide range of parameter values to investigate the structural stability and robustness of the numerical model.

**Modeling TNF activity on cells**

*A short review of TNF biology*

For the sake of clarity, we briefly review the basic facts on TNF. More detailed descriptions can be found in recent reviews such as (7).

TNF is a homotrimeric molecule that binds to two different receptors: TNF-R1 and TNF-R2 (7). TNF-R1 appears to be the key mediator of TNF signaling in both normal and tumor cells and for this reason we focus our analysis on this receptor. TNF-R1 has three subunits, and their cytoplasmic tails must be juxtaposd to trigger intracellular signaling. Two models of TNF-R1 subunits recruitment have been proposed over the years: the first assumes that the receptor subunits juxtapose upon binding of homotrimeric TNF which, therefore, would drive the assembly of active TNF-R1; the second, which is supported by recent experimental evidences, suggests that TNF-R1 subunits self-assemble in the absence of TNF thanks to a conserved extracellular domain called the Pre-Ligand Assembly Domain (PLAD, 17). In this case, signaling by pre-assembled receptors before TNF binding would be prevented by cytosolic negative regulators sych as the Silencer of Death Domain (SODD, 18).

Binding of TNF to TNF-R1 initiates a series of biochemical events in the cell that take place at the cytoplasmic tails of the receptor subunits and, in particular, at their specialized domains called Death Domains (DD). DD recruit the adaptor protein TRADD that acts as an assembly platform for at least two other proteins, RIP-1 and TRAF-2 (2, 3, 7). This multiproteic complex initiates the signaling cascades resulting in NF-κB activation and hence gene activation and cell survival (2, 3, 7). Among the genes that are expressed after NF-κB activation, there are those that code for the two proteins FLIP and IAP that inhibit the TNF apoptotic pathway (2, 3, 7).

It has been recently demonstrated that the TNF apoptotic pathway is initiated by TNF/TNF-R1 complexes internalized into endocytic vesicles (14). At this intracellular level, the multiproteic complexes associated to the receptors' tails modify and form the so-called Death Inducing Signaling Complex (DISC, 14), whereby TRADD recruits FADD and pro-caspase-8. This caspase then triggers the irreversible pathway leading to apoptosis and cell death. It has also been demonstrated that the fate of endosomes containing TNF/TNF-R1 complexes prior to their maturation into lysosomes is to fuse with vesicles from the trans-Golgi network (14). They may contain two inactivated enzymes, pro-A-SMase and pre-pro-CTSD, whose activation is also triggered by the multiproteic complexes formed at the TNF-R1 cytoplasmic tails upon the formation of the multivesicle structure, and in particular by active caspase-8. Activation of A-SMase/CTSD cascade is also capable of mediating apoptosis via Bid cleavage and caspase-9 activation (14).

Both the NF-κB and the apoptotic pathways comprise a series of complex intracellular reactions involving a number of enzymes and substrates (2, 3, 7). These have been the subject of intense modeling efforts aimed at explaining the response of individual cells to TNF from a systemic perspective at the molecular level (8-12). To the best of our knowledge these models do not focus on the different timing of the two pathways, but rather on the switching dynamics of the underlying biochemical system, resulting from feedback motifs detected at subcellular level. Indeed, current biological data indicate that activation of NF-κB and caspases occurs at different sites in the cells (at the cell membrane and upon internalization in endosome, respectively), at different times.

*Binding and internalization of TNF/TFN-R1 complexes*

Current biochemical data show that the TNF-R1 receptors rapidly self-trimerize at the cell membrane because of the PLAD domains and interact with TNF homotrimers (17). Thus, the mechanism of TNF binding to TNF-R1 can in principle be viewed as the result of the monomeric interactions between one molecule of TNF and one molecule of receptor. This semplification is futher supported by the following considerations:

1. the mechanism of receptor self-trimerization followed by ligand binding can be modeled by a set of 5 differential equations with 6 parameters. The model can fit experimental data of TNF binding, but one finds that the kinetics of receptor trimerization are much faster than the binding kinetics, and thus the trimerized receptor behaves as an effective monomer. In addition, experimental determinations of the parameter values for intermediate binding reactions are not available;

2. experimental data of TNF dose-dependent binding to cells or of TNF binding kinetics have already been succesfully fitted with mathematical models that do not take into consideration the trimerization of receptors, neither TNF-induced nor TNF-independent, but, on the contrary, by classical models based on the law of mass action between two molecules (19-21).

In a series of papers, Vuk-Pavlović and colleagues (19, 20) systematically investigated a number of models of TNF binding and internalization in cells and, by quantitative comparison with actual experimental data, they produced a satisfactory basic model. Since the model by Bajzer et al. (19) fitted experimental data well, we use it to describe the early events of TNF interactions with cells.

However the original model of Bajzer et al. (19) must be updated to account for some novel aspects of TNF biology. In particular, Bajzer et al. assumed that internalized ligand/receptor complexes could be recycled back at the cell surface (19). Recent data, however, show that the final fate of the endosomes containing TNF complexes is to maturate to lysosomes by progressive fusion with vesicles from the trans-Golgi network loaded with lysosomal enzymes (see also the previous section and ref. 14), and thus it is highly probable that TNF/TNF-R1 complexes do not recycle at all but are finally degraded into lysosomes. Therefore we modify the model by Bajzer et al. as follows (see also Fig.1):

$$\frac{d[R]}{dt} = V_r - k_d[R] - k_{on}[L][R] + k_{off}[Nc]$$
$$\frac{d[L]}{dt} = -k_{on}[L][R] + k_{off}[Nc]$$
$$\frac{d[Nc]}{dt} = k_{on}[L][R] - (k_{off} + k_{in})[Nc] \quad [1]$$
$$\frac{d[Nin]}{dt} = k_{in}[Nc] - k_{deg}[Nin]$$

where square brackets denote molar concentrations of free TNF-R1 receptors (*R*), free TNF (*L*), TNF/TNF-R1 complexes bound at the cell membrane (*Nc*) and internalized complexes (*Nin*). Here $k_{on}$ and $k_{off}$ are the association and dissociation rate constants for TNF binding to TNF-R1, respectively, $k_{in}$ is the internalization rate constant of TNF/TNF-R1 complexes and $k_{deg}$ is the rate constant of lysosomal degradation of the complexes.

The two parameters $V_r$ and $k_d$ where introduced by Bajzer et al. (19), although with a slightly different notation, to describe "the zero-order rate of insertion of receptors into the membrane and the turnover (internalization) rate constant of ligand-free receptors" (19, 20) respectively. This is an important aspect of the model since, in the absence of TNF, it reaches a steady concentration of receptors at the cell surface given by:

$$[R]_{[L]=0} = \frac{V_r}{k_d} \quad [2]$$

In addition, for long times these terms prevent receptor loss (i.e. downmodulation) from the cell surface, a process that would undesirably result in cell resistance to TNF independently of the dynamic interplay between the intracellular paths triggered by TNF.

*Modeling the intracellular signaling pathways triggered by TNF*

Fig.1 shows how our minimal model maps onto the main biochemical paths triggered by TNF binding to its receptors. The figure, and hence our modeling effort, has been inspired by the recent work of Schneider-Brachet et al. (14) that elegantly demonstrates that the pathway leading to NF-κB activation and cell survival is initiated at the cells surface upon the formation of TNF/TNF-R1 complexes, whereas the one that leads to apoptosis and cell death by internalized complexes. In addition to the basic observations by Schneider-Brachet et al., we have integrated this circuits by implementing the NF-κB-mediated transcription of genes coding for caspase-8 inhibitors such as FLIP. In this way, the two intracellular pathways interact dynamically, inasmuch as the cell survival pathway - that starts earlier since it does not require internalization of TNF/TNF-R1 complexes - can inhibit the apoptotic path.

Here we model both biochemical circuits by means of only two molecular species, that we denote with *B* and *A*, that collectively summarize the various reactions leading to cell survival and death, respectively, and the interplay between the two paths. The, molecules *B* and *A* can be loosely identified with NF-κB/FLIP and caspase-8, respectively (Fig.1). We assume that after the initial trigger both pathways proceed irreversibly to their endpoint. In this way we neglect many details of both pathways which involve a number of different molecular actors, and thus we neglect all those reactions that probabily serve to fine tune the effects of TNF. Finally, we introduce a population variable $f(t)$, the fraction of surviving cells at time $t$. The equations for $A$ and $B$ are:

$$\frac{d[B]}{dt} = \beta[Nc] - k_{Bdeg}[B]$$
$$\frac{d[A]}{dt} = \alpha[Nin] - \gamma[B][A] - k_{Adeg}[A] \qquad [3]$$
$$\frac{df(t)}{dt} = -\kappa[A]f(t)$$

where the variables $Nc$ and $Nin$ are the same as in the differential system [1].

We see from the differential system [3], that the cell survival signal, modeled phenomenologically by means of the chemical species $B$, depends on the number of TNF-TNF-R1 complexes at the cell surface (e.g. $Nc$), with rate constant parameter $\beta$. On the other hand, the apoptotic signal, modeled phenomenologically by means of the chemical species $A$, depends on the number of internalized ligand/receptor complexes (e.g. $Nin$) with rate constant $\alpha$. The cell survival pathway inhibits the apoptotic one by destroying $A$ with rate $\gamma[B]$. Finally, both $B$ and $A$ can be degraded by means of ubiquitination and proteasome cleavage and/or irreversibly inhibited by other molecular species, and these processes are described by the rate constants $k_{Bdeg}$ and $k_{Adeg}$, respectively.

The surviving fraction $f(t)$ has also a probabilistic meaning at the single-cell level, since $\kappa[A]dt$ is the (Poisson) probability that a single cell dies during the time interval $(t, t+dt)$. This function, therefore, models cell death as a single-hit mechanism with rate $\kappa[A]$, and this notion is supported by a number of different cytotoxicity experiments (see e.g. refs. 22-26). Notice also that if $[A]$ is kept constant we recover a familiar formula for the surviving fraction (22, 26):

$$f(\Delta t) = e^{(-\kappa [A] \Delta t)} \quad\quad\quad [4]$$

*Mathematical methods*

Eventually, the systems of equations [1] and [3] define a complete model of TNF bioactivity on target cells, and we have used the *Mathematica* programming environment to carry out all integrations, fits and robustness tests (Wolfram Research, Inc., 100 Trade Center Drive, Champaign, IL, U.S.A.).

**Results**

*Binding and internalization kinetics*

We have estimated the parameters related to TNF binding and internalization by fitting equations [1] to available experimental data. The choice of the appropriate data, however, is not trivial. Fitting of equations [1], in fact, requires *a priori* knowledge of:

1. number of cells used in the assays and number of receptor molecules per cell, in order to fix the initial value for the variable $[R]$;

2. effective free TNF concentration in the experiment (in general TNF is labelled with radioactive isotopes such as $^{125}$I, and it would be important to know if TNF has retained full biological activity after the labeling procedure);

3. specific activity of labeled TNF in order to convert e.g. radioactive counts to grams of TNF and finally to moles units;

4. reaction volume to convert data to concentration units;

5. fraction of internalized TNF molecules that undergo proteolysis.

The latter point is particularly important for experiments that last several hours, where the degradation kinetics is expected to contribute much to the degradation of internalized molecules. From an experimental point of view, this fraction can be estimated by measuring the amount of radioactivity that cannot be precipitated by, e.g., trichloroacetic acid (or $(NH_4)_2SO_4$) in cell extracts after removing bound TNF from cell membranes (e.g. by treating cells with mild acidic solutions at 4 ºC or by other means) (27). Finally, we take into account only experiments carried out at 37 ºC, where internalization proceeds normally, and not at 4 ºC.

After extensive search of the scientific literature we found that the experiments in (21) match most of the above conditions. Grell *et al.* (21) used HeLa cells that express only the isoform R1 of the TNF receptor at a reported amount of approximately 3000 molecules per cell. The number of cells used and the residual biological activity of labeled TNF were also given. Finally experiments were carried out at 37 ºC and lasted 15 minutes, thus allowing TNF molecules to internalize and only a small fraction thereof to reach the lysosomes where degradation can occur (28). Since Grell *et al.* (21) did not discriminate between bound and internalized TNF molecules, the sum $[Nc](t)+[Nin](t)$ was fitted to experimental data of total cell-associated radioactivity. Fig.1 shows the data in (21) along with the best fit to the data using the set of differential equations [1].

Although the data set contains only 9 data points and the model has 6 parameters, the fit can still be performed, (we find $\chi^2 = 9.0$, which corresponds to a 3% statistical significance), and the estimated parameter values are as follows:

$V_r = 1 \cdot 10^{-12} M \cdot min^{-1}$, $k_d = 0.1 \cdot min^{-1}$, $k_{on} = 6.5 \cdot 10^8 M^{-1} \cdot min^{-1}$, $k_{off} = 0.1 \cdot min^{-1}$, $k_{in} = 0.05 \cdot min^{-1}$ and $k_{deg} = 0.16 \cdot min^{-1}$. Unfortunately, the small number of available experimental data did not allow us to estimate the error bounds for the parameter values.

*Dose-response cytotoxicity assays*

Grell *et al.* (21) reported dose-response data with HeLa cells. However, cell viability was measured by spectrophotometry upon crystal violet staining of the cells and background values were not given. We therefore chose to compare model outputs with the data in Scherf *et al.* (29). These authors measured TNF cytotoxicity against MCF7 (human breast carcinoma cells) and Colo205 (human colon carcinoma cells) by the $^3$[H]-leucine incorporation assay, a method that provides very low background values. Unfortunately:

1. to the best of our knowledge the number of TNF receptors per cell is not known for these tumor cell lines;

2. dose-response cytotoxicity assays measure cell killing at different doses of a drug, and these are generally obtained by logarithmic serial dilutions (typically 10-fold) of TNF stocks. As a consequence, experimental data are limited and the goodness of fit cannot be statistically evaluated.

Presently we do not aim at precise parameter estimates for a specific cell line, but rather at a general test of the validity of the model; we also seek to determine its robustness with respect to parameter changes (see also the Discussion section).

Therefore we used equations [1] with the parameter values given above to model TNF binding and internalization and then attempted to fit cytotoxicity data in (29) using equations [3]. The initial conditions that are needed to solve the differential system were taken from the experiments by Scherf *et al.* (29), that were carried out by plating a different number of cells in a higher volume of growth medium rather than those by Grell *et al.* (21). The results are shown in Fig.3.

The estimated parameter values are as follows:

1. MCF7 cells: $\alpha = 0.071\ min^{-1}$, $\beta = 0.33\ min^{-1}$, $\gamma = 1.2 \cdot 10^{10}\ M^{-1} \cdot min^{-1}$, $k_{Adeg} = k_{Bdeg} = 0.033\ min^{-1}$, $\kappa = 6.1 \cdot 10^{9}\ M^{-1} \cdot min^{-1}$;

2. Colo205 cells: $\alpha = 0.071\ min^{-1}$, $\beta = 0.33\ min^{-1}$, $\gamma = 1.15 \cdot 10^{10}\ M^{-1} \cdot min^{-1}$, $k_{Adeg} = 0.033\ min^{-1}$, $k_{Bdeg} = 0.018\ min^{-1}$, $\kappa = 4.0 \cdot 10^{9}\ M^{-1} \cdot min^{-1}$.

To test the stability of the outputs in Fig.3 we simulated the cytotoxic effects for MCF7 cells at different TNF doses administered for increasing times. The results in Fig.4 show that the simulated cytotoxic effects saturates after approximately 44 hours, and that no further TNF-mediated cytotoxicity can be observed after that time. It is worth noting that the maximal observed cytotoxic effect in Fig.4 corresponds to a surviving fraction $f = 0.038$. That is to say, 3.8 % of the cells survive independently of TNF concentration and treatment time.

*Exploring the space of parameters*

Our model of TNF cytotoxicity is collectively given by equations [1], describing uptake and internalization of TNF, and equations [3], describing intracellular processes activated by TNF that lead to either cell survival or death. Parameter values in equations [1] have been estimated by fitting with a good level of statistical significance, and all the model assumptions can be subjected to experimental validation. Model equations [3], instead, describe phenomenologically the intricate web of molecular interactions that constitute the intracellular signal transduction machinery triggered by binding of TNF to TNF-R1. The paucity of experimental data (see e.g. ref. 30) makes it difficult to compare model outputs with specific experimental observations. In this case it is important to study whether the model can still provide reasonable outputs, or fail, for a broad range of parameter values.

Model equations [3] contain 5 independent parameters, namely $\alpha$, $\beta$, $\gamma$, $k_{Adeg}$ and $k_{Bdeg}$ that tune the intracellular activation and degradation of molecules *A* and *B* and their interactions. Parameter $\kappa$ models cell death as a single-hit mechanism depending on the concentration of molecule *A*, and thus the exponential behaviour of the function $f(t)$ (see equation [4]) is not altered by the different choice of $\kappa$ values that can only adjust the shape of the sigmoid curve that describes the surviving fraction. Thus, the space of the parameters is effectively 5-dimensional and here we explore several sections of this space as a function of TNF concentration.

Fig.5 shows the model outputs calculated for a broad range of parameter values as a function of TNF concentrations in the range $11^{-14}$-$11^{-9}M$. The output is the maximum fraction of surviving cells that can resist a given amount of TNF after a very long time treatment (i.e., $\lim_{t \to \infty} f(t) = f(\infty)$), and this was calculated after 20.000 min (approx. 14 days) of TNF treatment to allow any transient to settle down. In Fig.6 we also change the values of two parameters at the same time. In this case TNF concentration was set to $1.7 \cdot 10^{-11} M$, which is the IC50 for MCF7 cells (IC50 = concentration that inhibit 50% cell survival, see Fig.3).

Results in both Fig.5 and Fig.6 show that no unexpected and/or undesired patterns emerge, and that the model actually describes a balance between cell survival and death for a broad range of parameter values. Thus the model is structurally stable and robust with respect to parameter variations.

**Discussion**

We have developed a minimal quantitative model of TNF cytotoxicity. The model is minimal because it takes into consideration only those reactions that, in our opinion, are essential to describe the action of TNF on cell survival and death.

We have modeled TNF binding and its uptake by cells, and it is worth noting that the estimated parameter values are biologically relevant. In fact:

1. the ratio $\frac{V_r}{k_d}$ (equation [2]) determines the concentration of TNF receptors in the cell population at equilibrium. Substituting the estimated values, we obtain $[R]_{[L]=0} = 10^{-11} M$ and, considering the experiments by Grell *et al.* (21) where approximately $4.5 \cdot 10^5$ cells were used in a reaction volume of 150 µl, we find that each cells expresses on average approx. $2.26 \cdot 10^3$ TNF receptors (which corresponds to a measured concentration of $1.126 \cdot 10^{-11} M$). This value is in good agreement with experimental data collected with HeLa cells (21);

2. the dissociation constant at equilibrium is defined as $K_d = \frac{k_{off}}{k_{on}}$. Using the above values, $K_d = 1.5 \cdot 10^{-10} M^{-1}$ approximately 10 times higher than reported by Grell *et al.* (21) for HeLa cells. However, these authors estimated the dissociation constant by non-linear regression of experimental data with standard binding equations and they did not take into account internalization of bound molecules;

3. internalization and degradation rate constants show values that are in a reasonable range from a biological perspective. In fact they imply that, on average, the characteristic time of TNF internalization (at the considered concentration) is 20 min and that approximately 6 more minutes are required by internalized complexes to reach the lysosomes where degradation occurs (28).

Moreover the rate constant $\beta$ parameterizes the triggering kinetics of the survival signal that initiates at the cell membrane level upon binding of TNF to TNF-R1. From its estimated value, we can calculate that the survival signal is very quick and starts within a few minutes (approximately 3 min) after TNF binding. Using mutant cells expressing a TNF-R1 where the internalization domain - called TRID - was deleted (TNF-R1 ΔTRID), Schneider-Brachert *et al.* (14) clearly demonstrated that within 3 min after receptor activation, a complex of adaptor proteins (consisting of RIP-1 and TRAF-2) is formed at the cell surface, signaling for NF-κB activation and cell survival. However, the proteins TRADD, FADD and caspase-8 were not recruited, demonstrating the endocytosis and DISC formation are inseparable events (14). Thus, survival and death signals are temporarily separated and this is reflected in our model by the differences between the values of the parameters $\alpha$ and $\beta$.

We have extensively explored the parameter space of the model. Simulations show that the model has a stable behavior for a broad range of parameter values and that no unexpected patterns emerge (such as oscillations, chaos, etc.). Thus the model is structurally stable: obviously this is quite important and lends credibility to the model, because this indicates that different cells with unequal expression of key substrates and enzymes and/or showing a signal transduction network with a different topology have similar responses to TNF. This also means that our model is not specific for a given cell type and that it can be used to simulate the effects of TNF independently of the experimental settings of the original data on which the parameter estimate is based.

In our model there is no switching mechanism that selects cell survival or death signals, but rather a balance between the two pathways that produces partial cell killing even for long lasting and intense TNF treatments. The balance depends on environmental TNF concentration, and this observation might be important to explain cellular homeostasis during an immune response, i.e. the fine equilibrium between cellular activation and death (1). This equilibrium might further be balanced in favour of cell survival or death in real cells by the fine expression and/or degradation of intracellular molecular actors that transduce TNF signals (for example, in our model cell survival can be improved by increasing degradation of *A* and/or the expression of *B* molecules. See also ref. 5).

The model nicely fits cytotoxicity data with tumor cells. There is an increasing interest in modeling tumor growth and behaviors, and multi-scale models are now under development (15, 16, 31-34, and references cited therein). Our model of TNF cytotoxicity provides outputs in quantitative agreement with actual data, and it is robust to parameter changes. Moreover, a true fit of the model equations to experimental data is possible in this case because of the low number of model parameters, far fewer than those considered in current models that approach the TNF signal transduction pathways from a systemic perspective (e.g. > 50 parameters as in ref. 8). Obviously, parameter reduction also means that many details of the underlying biology must be neglected, but in this way the essential features of a biological mechanism - such as TNF action on cells - can be clearly investigated, captured, controlled and eventually implemented into more complex models of the cell. Further details of the TNF signal transduction machinery could then be added using the present model as a starting point.

Because of its favorable properties, our model is suitable for integration into complex multi-scale simulation programs of tumor growth such as VBL (15, 16). We plan to use the model to explore in detail the response of tumor cell clusters to TNF therapy and to investigate tumor/immune system interaction dynamics.

**Figure Legends**

Fig.1 - Modeling TNF cytotoxicity. **A.** Biological view of the main TNF paths (redrawn from 14). Binding of TNF to TNF-R1 recruits a molecular complex formed by TRADD, RIP-2 and TRAF-2, signaling for NF-κB activation, cell survival and expression of caspase inhibitors such as FLIP. During endocytosis, TRADD, FADD and caspase-8 are recruited to form the TNF-R1 associated DISC. Caspase-8 is autocatalytically cleaved during receptosome trafficking. **B.** Scheme of the biochemical paths that have been considered in the present model. TNF binds to TNF-R1 with association and dissociation rate constants given by $k_{on}$ and $k_{off}$ and the complexes are internalized with rate constant $k_{in}$. Binding activates signaling resulting in the formation of a molecule $B$ with rate constant $\beta$. Molecule $B$ inhibits the apoptotic signal which has been assumed to be proportional to the concentration of molecule $A$, and this process occurs with rate constant $\gamma$. The apoptotic signal $A$ is triggered by internalized TNF/TNF-R1 complexes with rate constant $\alpha$. Internalized receptors are finally degraded into lysosomes, and both molecular signals $A$ and $B$ are cleared off by proteolytic cleavage or by inhibitory binding with other molecules (rate constants $k_{deg}$, $k_{Adeg}$ and $k_{Bdeg}$, respectively). Equilibrium in the expression and downmodulation of TNF receptors have been assumed to occur at the cell surface (rate constants $V_r$ and $k_d$. See also ref. 19 and the text for details).

Fig.2 - Binding kinetics of TNF at 37 ºC. Symbols have been redrawn from (21) and refer to experiments carried out with HeLa cells. The curve represent the best fitting obtained with model equations [1].

Fig.3 - Fitting dose-response cytotoxicity data with the whole model. Data have been redrawn from (29) and refer to experiments carried out with MCF7 (black symbols) or Colo205 (grey symbols). The curves show the best fits of experimental data with equations [1] and [3]. Parameters in equations [1] assume the values estimated by fitting of binding kinetics data as shown in Fig.2. Experimental error was not given in the original data.

Fig.4 - Computing TNF cytotoxicity as a function of both TNF concentration and time. Simulations were carried out using the parameter values estimated by fitting cytotoxicity data collected with MCF7 cells. The fraction of surviving cells has been computed after 1, 2, 4, 6, 8, ..., 48 hours of treatment.

Fig.5 - Robustness of the model with respect to parameter changes. Key parameters describing the intracellular paths triggered by TNF binding to TNF-R1 were varied over a broad range of values, and model outputs have been computed at each indicated TNF concentrations. Outputs refer to the fraction of cells surviving treatment after 20.000 min ($f(\infty)$) and the corresponding values in panels **A-E** are given in gray scale colors. These can be compared to the reference bar given on top.

Fig.6 - Robustness of the model with respect to parameter changes. See the legend of Fig.5 and the text for details. Here, we varied the values of two parameters at the same time and explore different combinations of model parameters. The surviving fraction was computed for a fixed concentration of TNF that was set at the IC50 value measured for MCF7 cells (see Fig.3).

Figure 1

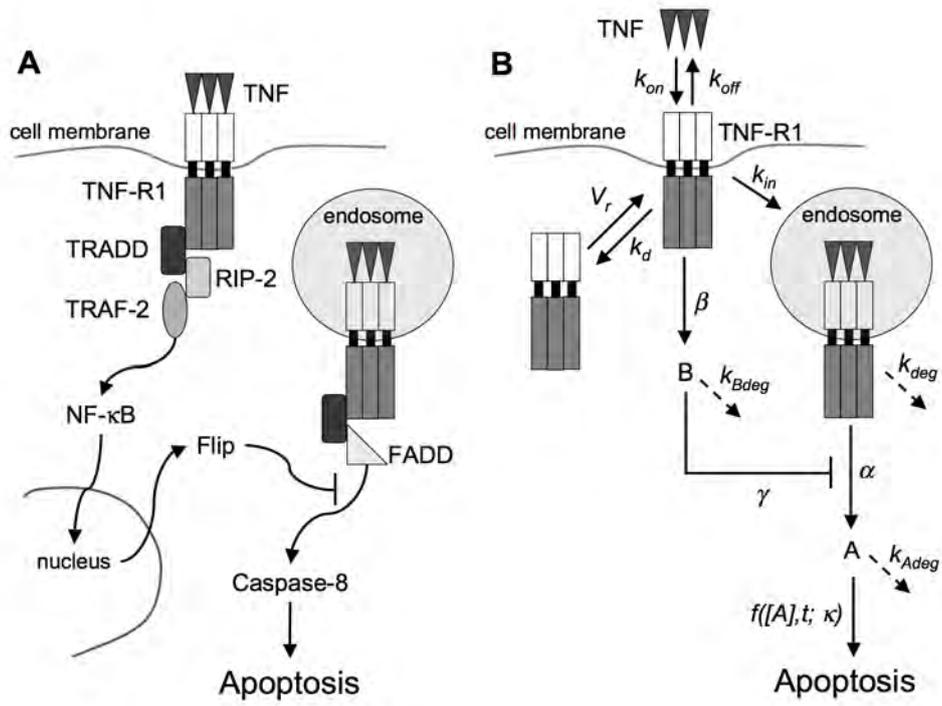

Figure 2

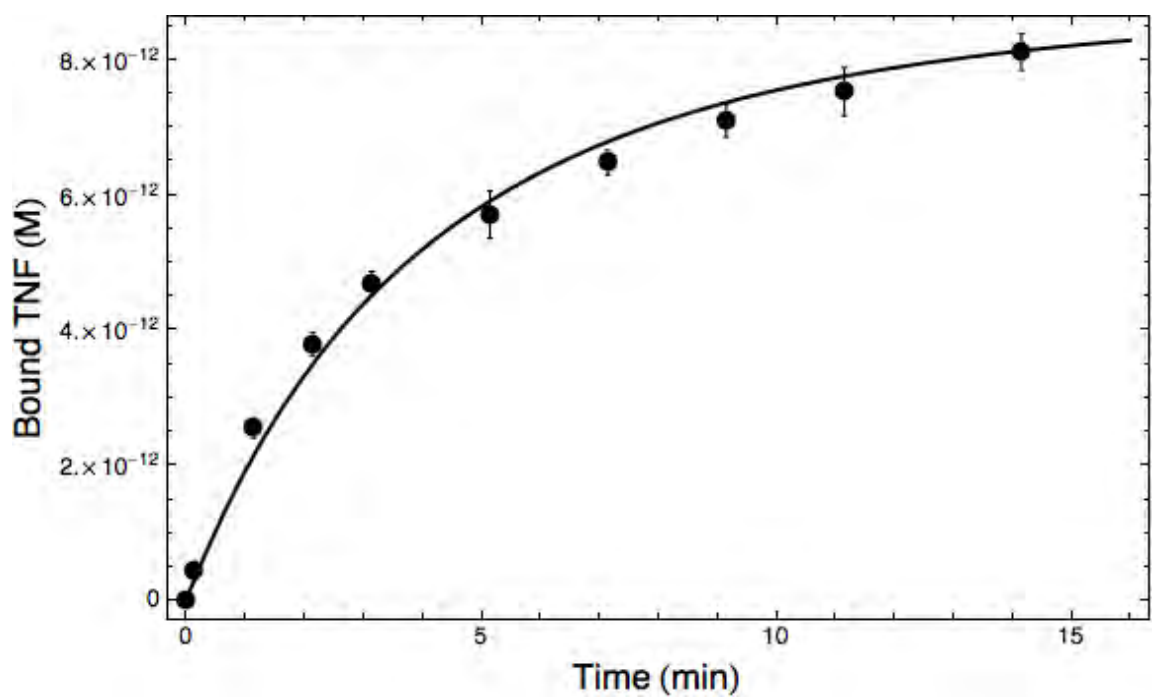

Figure 3

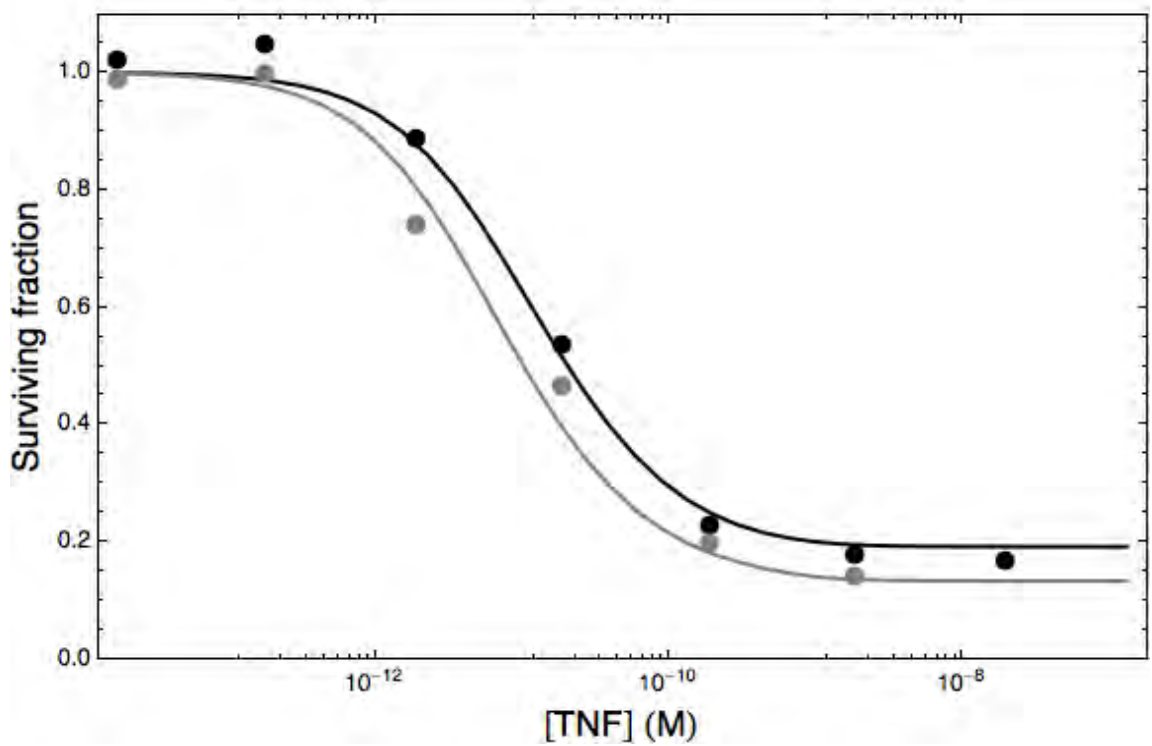

Figure 4

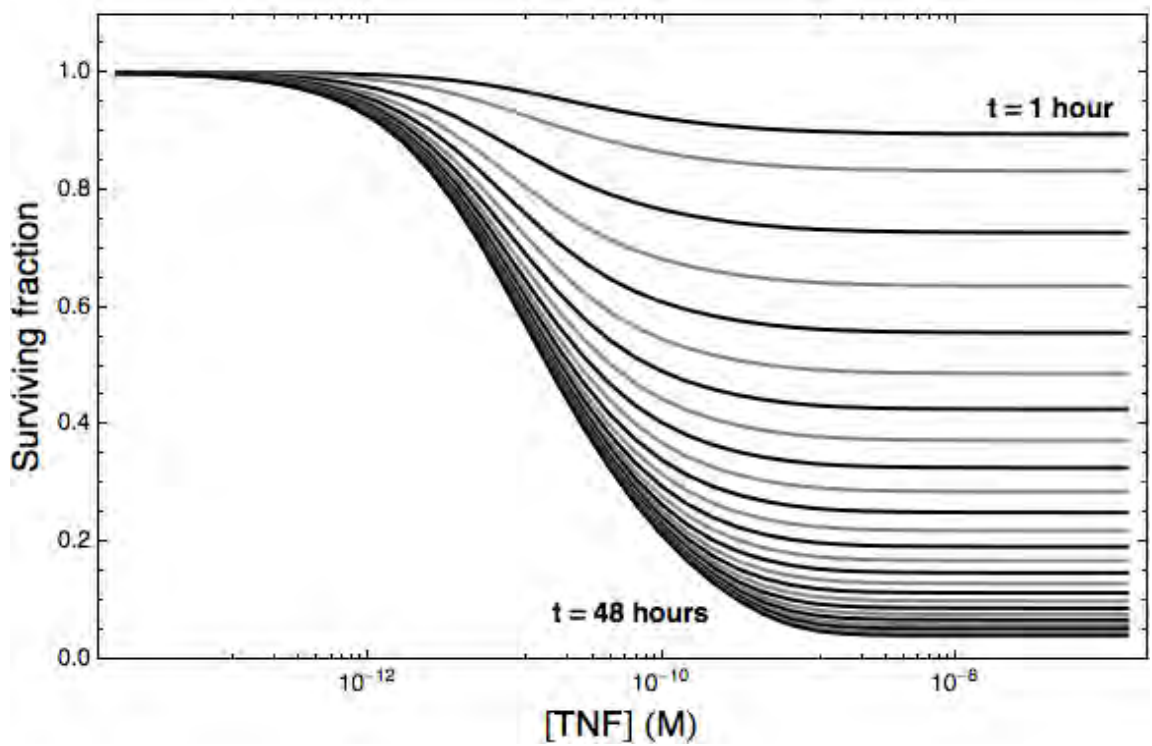

Figure 5

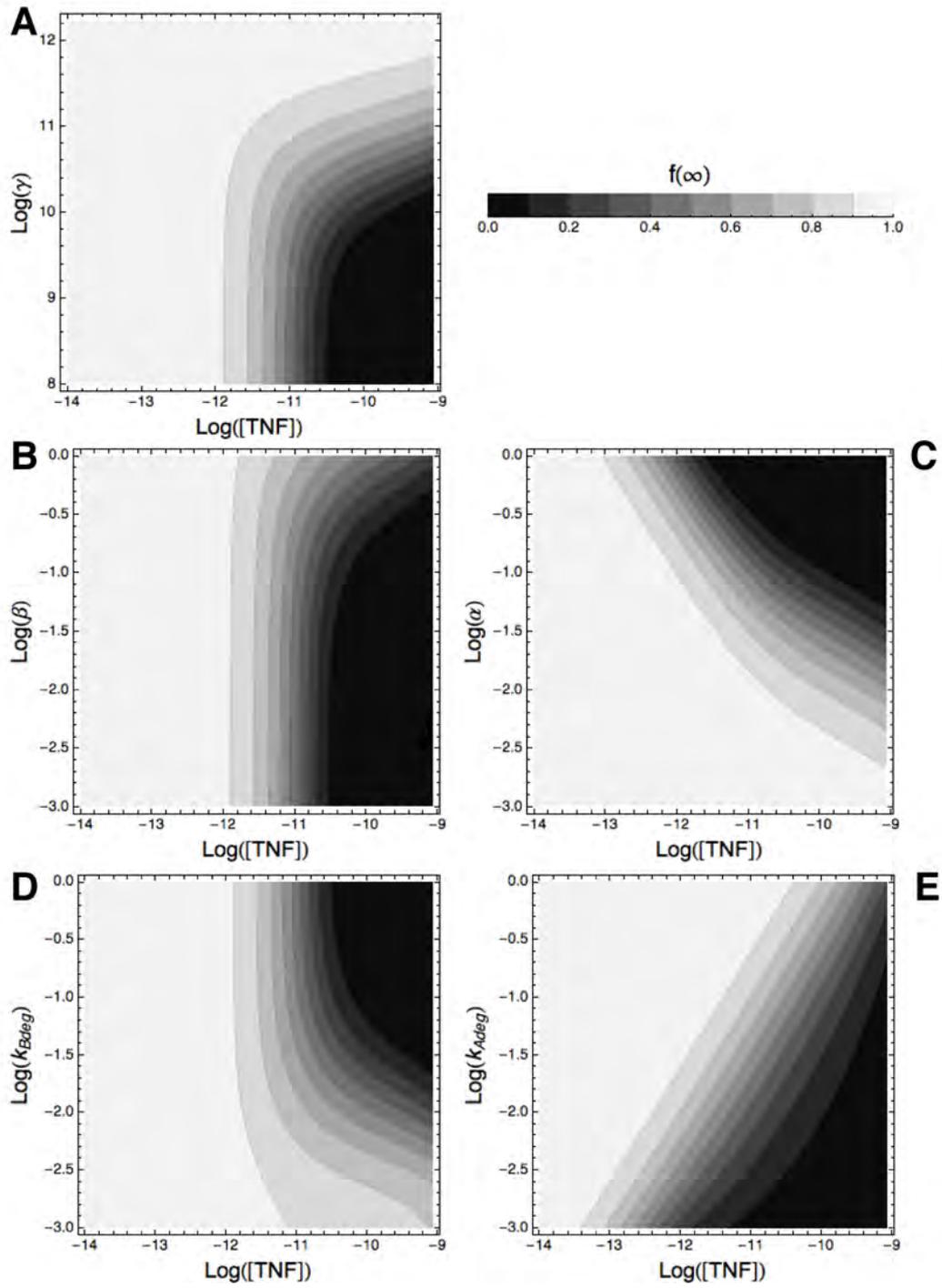

Figure 6

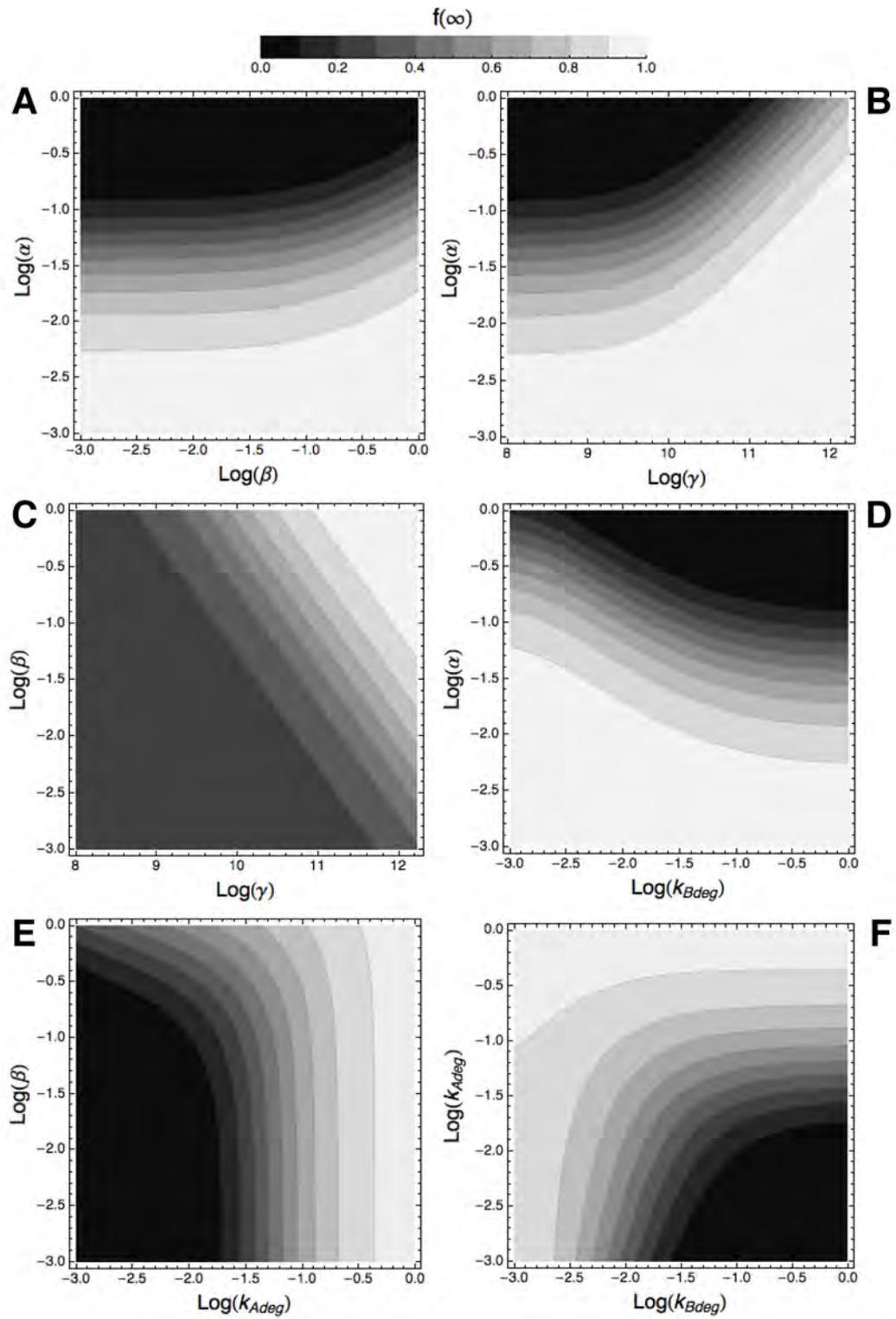